\documentclass[aps,pra,twocolumn,showpacs]{revtex4}
\usepackage{graphicx}
\usepackage{amsmath}
\usepackage{amsfonts}
\usepackage{amssymb}
\newcommand{\ket}[1]{|#1\rangle}
\newcommand{\bra}[1]{\langle #1|}
\newcommand{\Tr}{\text{Tr}}
\begin{document}
\today
\title{Non-Abelian quantum holonomy of hydrogen-like atoms}
\author{Vahid Azimi Mousolou\footnote{Electronic address: vahid.mousolou@lnu.se} and
Carlo M. Canali\footnote{Electronic address: carlo.canali@lnu.se}}
\affiliation{School of Computer Science, Physics and Mathematics, Linnaeus University,
Kalmar, Sweden}
\author{Erik Sj\"oqvist\footnote{Electronic address: erik.sjoqvist@kvac.uu.se}}
\affiliation{Department of Quantum Chemistry, Uppsala University, Box 518,
Se-751 20 Uppsala, Sweden}
\affiliation{Centre for Quantum Technologies, National University of Singapore,
3 Science Drive 2, 117543 Singapore, Singapore}
\begin{abstract}
We study the Uhlmann holonomy [Rep. Math. Phys. {\bf 24}, 229 (1986)] of quantum states
for hydrogen-like atoms where the intrinsic spin and orbital angular momentum are coupled
by the spin-orbit interaction and subject to a slowly varying magnetic field. We show
that the holonomy for the orbital angular momentum and spin subsystems is non-Abelian,
while the holonomy of the whole system is Abelian. Quantum entanglement in the states of
the whole system is crucially related to the non-Abelian gauge structure of the subsystems.
We analyze the phase of the Wilson loop variable associated with the Uhlmann holonomy,
and find a relation between the phase of the whole system with corresponding marginal
phases. Based on the result for the model system we provide evidence that the phase
of the Wilson loop variable and the mixed-state geometric phase [E. Sj\"oqvist {\it et al.} 
Phys. Rev. Lett. {\bf 85}, 2845 (2000)] are in general inequivalent.
\end{abstract}
\maketitle
\section{Introduction}
The pioneering work by Berry \cite{berry84} and Wilczek and Zee \cite{wilczek84},
have triggered considerable interest in effective gauge structures in the adiabatic
evolution of non-relativistic quantum systems. Non-Abelian quantum holonomies have
been examined in the context of nuclear rotations of diatoms \cite{moody86},
nuclear quadrupole resonance \cite{tycko87}, semiconductor heterostructures
\cite{arovas98}, trapped atoms \cite{unanyan99}, quantum optics \cite{pachos00},
and superconducting systems \cite{faoro03}. It has been pointed out
\cite{zanardi99} that non-Abelian holonomy may be used in the construction of
universal sets of quantum gates for the purpose to achieve
fault tolerant quantum computation.

In Refs. \cite{wilczek84,moody86,tycko87,arovas98,unanyan99,pachos00,faoro03,zanardi99},
non-Abelian holonomies are related to the existence of degenerate energy eigenstates
that can be controlled by a set of slowly changing parameters. In contrast, Uhlmann
\cite{uhlmann86} has shown that non-Abelian gauge structures may appear along sequences
of density operators representing mixtures of quantum states, irrespective of the degeneracy
structure of the underlying Hamiltonian. Such non-Abelian structures may arise for
subsystems of composite systems undergoing adiabatic evolution, since the
marginal states are mixed if the instantaneous energy eigenstates of the whole system are
entangled.

The purpose of the present paper is to follow the Uhlmann approach to examine non-Abelian
gauge structure in the case of spin-orbit ($LS$) coupled hydrogen-like atoms subject to a
slowly varying magnetic field. We show that the adiabatic Uhlmann holonomies for the spin
($S$) and orbital ($L$) parts become non-Abelian although the one of the whole $LS$ state
is Abelian. Studies of the same model were carried out previously \cite{sjoqvist05} (see
also \cite{oh08}) by using the mixed-state geometric phase approach  developed in  Ref.
\cite{sjoqvist00a}. In particular, in Ref. \cite{sjoqvist05} it was shown that the mixed-state
geometric phases of the $L$ and $S$ subsystems always sum up to the standard pure state geometric
phase of the whole system. In contrast, we show  here that the phases of Wilson loop variables
associated with the Uhlmann holonomies satisfy this sum rule only for specific paths,
while for other paths there is a deviation of $\pi$ from the sum rule. This deviation
from the sum rule demonstrates a striking non-trivial difference between the Uhlmann holonomy
\cite{uhlmann86} and the mixed-state geometric phase \cite{sjoqvist00a}.

The outline of the paper is as follows. In Sec.~\ref{sec:uhlmann} the Uhlmann
holonomy is briefly reviewed. The corresponding parallel transport equations for
adiabatic rotation of angular momentum states are derived in Sec.~\ref{sec:uhlmannS}.
In Sec.~\ref{sec:hydrogen}, we compute the Uhlmann holonomies for the $L$ and $S$
subsystems as well as that of the total angular momentum. We examine in particular the
non-Abelian nature of the subsystem holonomies as well as the additivity of the
phases of the Wilson loop variables associated with the Uhlmann holonomies. The
paper ends with the conclusions.

\section{Uhlmann holonomy}
\label{sec:uhlmann}
In this Section we summarize the main definitions and properties of the Uhlmann holonomy
\cite{uhlmann86}. Let $C: [0,1] \ni t \rightarrow \rho_t$ be a smooth path of density
operators acting on some Hilbert space $\mathcal{H}$. An operator $W_t$ such that
$\rho_t = W_t W_t^{\dagger}$ is called an amplitude of $\rho_t$. $W_t$ can be written
as $W_t = \sqrt{\rho_t} V_t$, where the ``phase factor'' $V_t$ is a partial isometry
\cite{remark1} on $\mathcal{H}$. For any choice of $W_0 \equiv \widetilde{W}_0$, there
is a differentiable path $[0,1] \ni t \rightarrow \widetilde{W_t}$ of amplitudes over
$C$ that satisfy the parallel transport condition
\begin{eqnarray}
\widetilde{W}^{\dagger} d \widetilde{W}  =
d \widetilde{W}^{\dagger} \widetilde{W} ,
\label{paralleltransport}
\end{eqnarray}
where $d \widetilde{W} = dt \dot{\widetilde{W}}_t$. Inserting $\widetilde{W} =
\sqrt{\rho} \widetilde{V}$ into Eq. (\ref{paralleltransport}) yields
\cite{hubner93}
\begin{eqnarray}
d \widetilde{V} \widetilde{V}^{\dagger} \rho +
\rho d \widetilde{V} \widetilde{V}^{\dagger} =
d \sqrt{\rho} \sqrt{\rho} - \sqrt{\rho} d \sqrt{\rho} .
\label{eq:pt}
\end{eqnarray}
By solving for $\widetilde{V}$, we define the Uhlmann holonomy
associated with the path $C$ to be
\begin{eqnarray}
U_{{\textrm{uhl}}} \boldsymbol{(} C \boldsymbol{)} =
\widetilde{V}_1 \widetilde{V}_0^{\dagger} .
\label{Uh}
\end{eqnarray}
The operator $U_{{\textrm{uhl}}} \boldsymbol{(} C \boldsymbol{)}$ is a unique
partial isometry (unique unitary if all $\rho_t$ are full rank), gauge invariant
(i.e., independent of the choice of $\widetilde{W}_0$), and reparametrization invariant
(i.e., independent of the speed of the evolution along $C$); thus, $U_{{\textrm{uhl}}}
\boldsymbol{(} C \boldsymbol{)}$ is a property of the path $C$.

In Uhlmann's approach, a system in a mixed state is thought to
be a subsystem embedded in a larger quantum system which is in a pure state. The pure
state is referred to as a purification of the mixed state. This is accomplished by introducing
an auxiliary system with which the original system is entangled. The purification is
equivalent to the amplitude $W_t$ of $\rho_t$. If $\widetilde{W}_t$ satisfies the condition
in Eq. (\ref{paralleltransport}) along the path $C$, then, inspired by the pure state
geometric phase \cite{mukunda93}, one may assign the phase
\begin{eqnarray}
\varphi_{\textrm{uhl}} = \arg \left[ \Tr (\widetilde{W}_0^{\dagger} \widetilde{W}_1)  \right]
\label{Uhlmannphase0}
\end{eqnarray}
to the Uhlmann holonomy $U_{{\textrm{uhl}}} \boldsymbol{(} C \boldsymbol{)}$. The definition
$\varphi_{\textrm{uhl}}$ for Uhlmann phase has been used to investigate theoretically
\cite{slater02,ericsson03,rezakhani06} and experimentally \cite{du07} a possible relationship
between the Uhlmann holonomy \cite{uhlmann86} and the mixed-state geometric phase $\beta$
\cite{sjoqvist00a}. The latter differs considerably from $\varphi_{\textrm{uhl}}$ in that
it is, for cyclic unitary evolution, the sum of geometric phase factors of the eigenstates
of the density operator weighted by the corresponding eigenvalues. Thus, contrary to
$\varphi_{\textrm{uhl}}$, $\beta$ is essentially a decomposition dependent and Abelian
geometric phase concept.

On the other hand, as for the case of the non-Abelian Wilczek-Zee phase factor \cite{wilczek84},
the Uhlmann holonomy in Eq. (\ref{Uh}) for cyclic evolutions takes the form of a Wilson loop
${\bf P} e^{-i \oint_{C} A}$ for a vector potential $A$, where ${\bf P}$ stands for path
ordering. On the basis of this fact and
on the fact that the Wilczek-Zee phase factor is a natural extension of the Berry phase
\cite{berry84} to systems with degenerate spectra, one can argue that the phase of the
Wilson loop variable $\Tr \left( {\bf P} e^{-i \oint_{C} A} \right)$ is more natural
quantity than $\varphi_{\textrm{uhl}}$, and define the Uhlmann phase as
\begin{eqnarray}
\gamma = \arg\left[\Tr\left( U_{\textrm{uhl}}(C)\right)\right].
\label{Uhlmannphase}
\end{eqnarray}
Note that the two phase quantities $\varphi_{\textrm{uhl}}$ and $\gamma$ need not be
equal. The focus of this paper is to turn our attention to the Uhlmann phase defined in
Eq. (\ref{Uhlmannphase}). In particular, we demonstrate that $\gamma$ behaves very
differently from the mixed-state geometric phase \cite{sjoqvist00a}

\section{Uhlmann holonomy of angular momenta}
\label{sec:uhlmannS}
As an introduction to the model studied in Sec.~IV, we consider the case of adiabatic
transport of a quantum angular momentum ${\bf S}$. This angular momentum is assumed
to be coupled to another quantum angular momentum ${\bf S}^{(r)}$. The coupling is
assumed to be spherically symmetric. Both ${\bf S}$ and ${\bf S}^{(r)}$ are
exposed to an external classical magnetic field ${\bf B}$. The Hamiltonian of the
system takes the form $H = U^{{\textrm{tot}}} (\theta,\phi) H_z {U^{{\textrm{tot}}}}^{\dagger}
(\theta,\phi)$, where $U^{{\textrm{tot}}}(\theta,\phi) = e^{-i\phi S_z^{{\textrm{tot}}}}
e^{-i\theta S_y^{{\textrm{tot}}}}e^{i\phi S_z^{{\textrm{tot}}}}$ and ${\bf S}^{{\textrm{tot}}} =
{\bf S} + {\bf S}^{(r)} = (S_x^{{\textrm{tot}}} , S_y^{{\textrm{tot}}} ,
S_z^{{\textrm{tot}}})$ ($\hbar = 1$ from now on). $\theta,\phi$ are spherical polar
angles parametrizing the direction ${\bf n} = {\bf B} / |{\bf B}|$ of the external
magnetic field.

$H_z$ is independent of the polar angles $(\theta,\phi)$ of the two-dimensional
parameter sphere $\mathcal{S}^{2}$ of all possible magnetic field directions.
We assume that $[S^{tot}_z,H_z]=0$ to make sure that $H$ is well defined on $\mathcal{S}^2$.
The energy eigenstates can be represented by the smooth vector-valued functions
$\ket{\psi^{(n)} (\theta,\phi)} = U^{{\textrm{tot}}} (\theta,\phi) \ket{\psi_z^{(n)}}$,
well defined on the open patch $\mathcal{S}^2 - \{ \theta = \pi \}$, and $\ket{\psi^{(n)}
(\theta,\phi)}' = \bar{U}^{{\textrm{tot}}} (\theta,\phi) \ket{\psi_z^{(n)}} =
U^{{\textrm{tot}}} (\theta,\phi)e^{-2i\phi S^{tot}_z} \ket{\psi_z^{(n)}}$, well defined
on the open patch $\mathcal{S}^2-\{ \theta = 0 \}$. Here, $H_z \ket{\psi_z^{(n)}} =
E^{(n)}\ket{\psi_z^{(n)}}$. The vectors $\ket{\psi^{(n)} (\theta,\phi)}$ and $\ket{\psi^{(n)}
(\theta,\phi)}'$ define two monopole sections \cite{wu75} over the parameter sphere. These
sections are related by a single-valued gauge transformation so that
\begin{eqnarray}
\ket{\psi^{(n)} (\theta,\phi)}\bra{\psi^{(n)} (\theta,\phi)} =
\ket{\psi^{(n)} (\theta,\phi)}'\ '\bra{\psi^{(n)} (\theta,\phi)}
\end{eqnarray}
in any overlapping region on the parameter sphere.

The reduced density operator $\rho^{(n)} (\theta,\phi)$ representing the marginal state
of ${\bf S}$, corresponding to the $n$th energy eigenstate of $H$, is obtained by partial
trace $\Tr_r$ over the degrees of freedom associated with ${\bf S}^{(r)}$, i.e.,
\begin{eqnarray}
\rho^{(n)}(\theta,\phi) & = & U(\theta,\phi) \rho_z^{(n)} U^{\dagger} (\theta,\phi)
\nonumber \\
 & = & \bar{U}(\theta,\phi) \rho_z^{(n)} \bar{U}^{\dagger} (\theta,\phi) .
\end{eqnarray}
Here, $U(\theta,\phi) = e^{-i\phi S_z} e^{-i\theta S_y} e^{i\phi S_z}$ is the rotation
operator, $ \bar{U}(\theta,\phi)=U(\theta,\phi)e^{-2i\phi S_z}$, and $\rho_z^{(n)} =
\Tr_r \ket{\psi_z^{(n)}} \bra{\psi_z^{(n)}}$ is a ``reference'' state. $\rho^{(n)}
(\theta,\phi)$ defines the mixed state of our subsystem, for the $n$th energy eigenstate.

In the adiabatic regime, the path $\Gamma: [0,1] \ni t \rightarrow (\theta_t,\phi_t)$
on the parameter sphere $\mathcal{S}^2$ of magnetic field directions maps to the path
$C^{(n)}: [0,1] \ni t \rightarrow \rho^{(n)} (\theta_t,\phi_t)$ in state space of the
considered angular momentum. Let $\widetilde{V}^{(n)} = U(\theta,\phi) V^{(n)}$ be a
partial isometry that satisfies parallel transport along $\Gamma$. With $d = d\theta
\partial_{\theta} + d\phi \partial_{\phi}$, we have
\begin{eqnarray}
& dV^{(n)} {V^{(n)}}^{\dagger} \rho_z^{(n)} +
\rho_z^{(n)} dV^{(n)} V^{(n)\dagger}
\nonumber \\
 & = -2i\sqrt{\rho_z^{(n)}} \left[ d\phi (1-\cos \theta) S_z \right.
\nonumber \\
 & + d\phi \sin \theta (S_x \cos \phi + S_y \sin \phi)
\nonumber \\
 & \left. - d\theta (-S_x \sin \phi + S_y \cos \phi) \right]
\sqrt{\rho_z^{(n)}}
\label{eq:ptspherical}
\end{eqnarray}
along $\Gamma$.

Repeating the calculation for the other monopole section, by using the
decomposition $\widetilde{V}^{(n)} = \bar{U}(\theta,\phi)\bar{V}^{(n)}$, leads to the equation
\begin{eqnarray}
& d\overline{V}^{(n)} {\overline{V}^{(n)}}^{\dagger}
\rho_z^{(n)} + \rho_z^{(n)} d\overline{V}^{(n)}
\overline{V}^{(n)\dagger}
\nonumber \\
 & = 2i\sqrt{\rho_z^{(n)}} \left[ d\phi (1+\cos \theta) S_z \right.
\nonumber \\
 & - d\phi \sin \theta (S_x \cos \phi - S_y \sin \phi)
\nonumber \\
 & \left. + d\theta (S_x \sin \phi + S_y \cos \phi) \right]
\sqrt{\rho_z^{(n)}}
\label{eq:altptspherical}
\end{eqnarray}
along $\Gamma$. Since $[\rho^{(n)}_z,S_z]=0$, $\overline{V}^{(n)}=e^{2i\phi S_z}V^{(n)}$ for the
choice $\overline{V}^{(n)}_0=e^{2i\phi_{0} S_z}V^{(n)}_0$ satisfies
Eq. \ref{eq:altptspherical}. Thus, the difference between $\overline{V}^{(n)}$
and $V^{(n)}$ precisely compensates the difference between the rotation operators
$U(\theta,\phi)$ and $\overline{U}(\theta,\phi)$ so that the Uhlmann holonomy
remains the same in the two representations. In other words, $U_{{\textrm{uhl}}}
\boldsymbol{(} C^{(n)}  \boldsymbol{)}$ is independent of which monopole
section we use. This implies that either of the above pair of monopole sections can be
used to calculate the Uhlmann holonomy for any path on the parameter sphere.

\section{Uhlmann holonomy of hydrogen-like atoms}
\label{sec:hydrogen}
In Ref. \cite{sjoqvist05}, the adiabatic geometric phases of the $LS$-coupled
hydrogen atom in a slowly rotating magnetic field ${\bf B} = B{\bf n} =
B(\sin \theta \cos \phi,\sin \theta \sin \phi,\cos \theta)$ were analyzed. The
adiabatic geometric phases of the whole system and of the orbital ($L$) and
spin ($S$) angular momentum subsystems were computed. In particular, it was
demonstrated that the subsystem phases add up to the phase of the whole system.
The purpose here is to compute the corresponding Uhlmann holonomies and to examine
their relation.

\subsection{Model system}
We consider the spin-orbit ($LS$) part
\begin{eqnarray}
H_{\bf n} & = &
g {\bf n} \cdot ({\bf L} + 2{\bf S}) +
2{\bf L} \cdot {\bf S}
\nonumber \\
 & = & U_J (\theta,\phi) H_z
U_J^{\dagger} (\theta,\phi) ,
\label{eq:soham}
\end{eqnarray}
of hydrogen-like atoms exposed to an external magnetic field pointing in a
direction defined by the unit vector ${\bf n}$. The first and second terms are
the Zeeman and $LS$-coupling contributions, respectively, $g$ is the Zeeman-$LS$
strength ratio (assumed to be time-independent), and ${\bf n}$ defines the adiabatic
parameter sphere $\mathcal{S}^2$ under slow changes in the direction of the external
magnetic field. We may choose
\begin{eqnarray}
U_X (\theta,\phi) & = &
e^{-i\phi X_z} e^{-i\theta X_y} e^{i\phi X_z} , \ X=L,S,J ,
\end{eqnarray}
where ${\bf J} = {\bf L} + {\bf S}$ is the total angular momentum and $H_z$ is
the Hamiltonian at the north pole ${\bf n}=(0,0,1)$ of the parameter sphere.

The Hamiltonian $H_z$ is block-diagonalizable in one- and two-dimensional
blocks with respect to the product basis with elements $\ket{l,m}\ket{\frac{1}{2},
\pm \frac{1}{2}} \equiv \ket{l,m} \ket{\pm}$ being the common
eigenvectors of ${\bf L}^2,L_z,{\bf S}^2,S_z$. Each block may be
labeled by the eigenvalue $\mu=-l-\frac{1}{2},-l+\frac{1}{2},
\ldots, l+\frac{1}{2}$ of $J_z$. The two extremal subspaces characterized
by $|\mu|=l+\frac{1}{2} \equiv \mu_{\textrm{\scriptsize{e}}}$ are
one-dimensional corresponding to the two product vectors
$\ket{\psi_{\pm}^{(l,\pm \mu_{\textrm{\scriptsize{e}}}})} = \ket{l,\pm l} \ket{\pm}$.
The remaining blocks are two-dimensional, each of which spanned by the vectors
$\ket{l,m=\mu-\frac{1}{2}} \ket{+}, \ket{l,m=\mu+\frac{1}{2}} \ket{-}$,
$|\mu| < l+\frac{1}{2}$. For each $\mu$, the corresponding energy eigenvectors
$\ket{\psi_{\pm}^{(l,\mu)}}$ become $LS$ entangled. The amount of $LS$ entanglement
may be measured in terms of concurrence \cite{wootters98}
\begin{eqnarray}
\mathcal{C}^{(l,\mu)}=\mathcal{C}(\psi_{\pm}^{(l,\mu)}) =
\left| \sin \alpha^{(l,\mu)} \right| ,
\end{eqnarray}
where
\begin{eqnarray}
\cos \alpha^{(l,\mu)} =
\frac{2\mu + g}{\sqrt{g^2 + 4g \mu + \big( 2l+1 \big)^2}} .
\end{eqnarray}
Note that $\mathcal{C}^{(l,\mu)}$ is independent of
the $\pm$ label and varies between $0$ for product states ($\alpha^{(l,\mu)} = 0,\pi$)
and $1$ for maximally entangled states ($\alpha^{(l,\mu)} = \pi/2$).

The instantaneous energy eigenvectors of $H_{\bf n}$ are related to the
eigenvectors of $H_z$ according to
\begin{eqnarray}
\ket{\psi_{\pm}^{(l,\mu)};\theta,\phi} & = &
U_J(\theta,\phi) \ket{\psi_{\pm}^{(l,\mu)}}
\nonumber \\
 & = & U_L(\theta,\phi) U_S(\theta,\phi) \ket{\psi_{\pm}^{(l,\mu)}} .
\end{eqnarray}
Let $\Gamma: [0,1] \ni t \rightarrow (\theta_t,\phi_t)$ be a parametrized path
on the parameter sphere. Assume that the external magnetic field slowly traverses
$\Gamma$ so that the adiabatic approximation is valid. With this assumption
$\Gamma$ maps to the paths
\begin{eqnarray}
 & & C^{(l,\mu)}_{X,\pm}: [0,1] \ni t \rightarrow \rho^{(l,\mu)}_{X,\pm} (\theta_t,\phi_t)
\nonumber \\
 & & = U_X(\theta_t,\phi_t) \rho^{(l,\mu)}_{X,\pm} U_X^{\dagger} (\theta_t,\phi_t), \ X=L,S,J,
\end{eqnarray}
in the spaces of density operators, where
\begin{eqnarray}
\rho^{(l,\mu)}_{J,\pm} & = & \ket{\psi_{\pm}^{(l,\mu)}}\bra{\psi_{\pm}^{(l,\mu)}} ,
\nonumber \\
\rho^{(l,\mu)}_{L,\pm} & = & \Tr_S \ket{\psi_{\pm}^{(l,\mu)}}\bra{\psi_{\pm}^{(l,\mu)}} ,
\nonumber \\
\rho^{(l,\mu)}_{S,\pm} & = & \Tr_L \ket{\psi_{\pm}^{(l,\mu)}}\bra{\psi_{\pm}^{(l,\mu)}} .
\end{eqnarray}
In the following subsections we compute the Uhlmann holonomies of the total
angular momentum $J$ of the atom and its subsystems $L$ and $S$ under the assumption
$g\neq0$.

\subsection{Holonomy of total angular momentum}
\label{subsec:total}
Let $V^{(l,\mu)}_{J,\pm}$ denote the solution of Eq. (\ref{eq:ptspherical}) with
reference state $\ket{\psi_{\pm}^{(l,\mu)}} \bra{\psi_{\pm}^{(l,\mu)}}$.
We obtain the partial isometry
\begin{eqnarray}
V^{(l,\mu)}_{J,\pm;1} = e^{-i\mu\int_{\Gamma} (1-\cos\theta) d\phi}
\ket{\psi_{\pm}^{(l,\mu)}} \bra{\psi_{\pm}^{(l,\mu)}} V^{(l,\mu)}_{J,\pm;0} ,
\end{eqnarray}
which yields the Uhlmann holonomy
\begin{eqnarray}
& U_{\textrm{uhl}} \boldsymbol{(} C^{(l,\mu)}_{J,\pm}\boldsymbol{)} =
e^{-i\mu \int_{\Gamma} (1-\cos\theta) d\phi}
\nonumber \\
 & \times U_J(\theta_1,\phi_1) \ket{\psi_{\pm}^{(l,\mu)}} \bra{\psi_{\pm}^{(l,\mu)}}
U_J^{\dagger} (\theta_0,\phi_0) .
\label{eq:1nmruhl}
\end{eqnarray}
To compare with the corresponding geometric phase factor $e^{i\beta \boldsymbol{(}
C^{(l,\mu)}_{J,\pm}\boldsymbol{)}}$, we note that while this geometric phase factor is a unit
modulus complex number, $U_{\textrm{uhl}} \boldsymbol{(} C^{(l,\mu)}_{J,\pm}\boldsymbol{)}$
in Eq. (\ref{eq:1nmruhl}) is a partial isometry with
${U_{\textrm{uhl}}^{\dagger}} \boldsymbol{(} C^{(l,\mu)}_{J,\pm}\boldsymbol{)}
U_{\textrm{uhl}} \boldsymbol{(} C^{(l,\mu)}_{J,\pm}\boldsymbol{)}$ and
$U_{\textrm{uhl}} \boldsymbol{(} C^{(l,\mu)}_{J,\pm}\boldsymbol{)}
{{U_{\textrm{uhl}}^{\dagger}} \boldsymbol{(} C^{(l,\mu)}_{J,\pm}\boldsymbol{)}}$
being projection operators onto the initial and final states,
respectively. In particular, this shows  that while the geometric phase
$\beta \boldsymbol{(} C^{(l,\mu)}_{J,\pm}\boldsymbol{)}$ is undefined if the two end
points of $C^{(l,\mu)}_{J,\pm}$ correspond to orthogonal states,
$U_{\textrm{uhl}} \boldsymbol{(} C^{(l,\mu)}_{J,\pm}\boldsymbol{)}$ is
a well-defined partial isometry. On the other hand, a direct calculation yields
\begin{eqnarray}
\Tr [U_{\textrm{uhl}} \boldsymbol{(} C^{(l,\mu)}_{J,\pm}\boldsymbol{)}] =
\big|\Tr [ U_{\textrm{uhl}} \boldsymbol{(} C^{(l,\mu)}_{J,\pm}\boldsymbol{)}]
\big| e^{i\beta \boldsymbol{(} C^{(l,\mu)}_{J,\pm}\boldsymbol{)}}
\end{eqnarray}
that demonstrates an explicit relation between the Wilson loop variable associated
with the Uhlmann holonomy and the geometric phase factor, unless $\Tr [U_{\textrm{uhl}}
\boldsymbol{(} C^{(l,\mu)}_{J,\pm} \boldsymbol{)}]$ vanishes, which happens precisely
when the initial and final states are orthogonal. These results establish a one-to-one
relation between the standard pure state geometric phase \cite{mukunda93} and the
corresponding Uhlmann holonomy of the whole system.

\subsection{Holonomy of the $L$ and $S$ subsystems}
\label{Subsystem holonomies}
Now we compute the Uhlmann holonomies of the $L$ and $S$ subsystems.
Let us start with the extremal states $\mu = \pm \mu_{\textrm{\scriptsize{e}}}$.
We note that the eigenvectors
\begin{eqnarray}
\ket{\psi_{\pm}^{(l,\pm \mu_{\textrm{\scriptsize{e}}})}; \theta,\phi} & = &
U_L (\theta,\phi) \ket{l,\pm l} U_S (\theta,\phi)
\ket{\pm}
\end{eqnarray}
of $H_{\bf n}$ are tensor products of states of the two subsystems $L$ and $S$. We
thus find the Uhlmann holonomies for the $L$ and $S$ subsystems as
\begin{eqnarray}
U_{\textrm{uhl}} \boldsymbol{(} C^{(l,\pm \mu_{\textrm{\scriptsize{e}}})}_{L,\pm}\boldsymbol{)}
 & = & e^{\mp i l \int_{\Gamma}(1-\cos \theta) d\phi}
\nonumber \\
 & & \times U_L (\theta_1,\phi_1) \ket{l,\pm l} \bra{l,\pm l}
U_L^{\dagger} (\theta_0,\phi_0) ,
\nonumber \\
U_{\textrm{uhl}} \boldsymbol{(} C^{(l,\pm \mu_{\textrm{\scriptsize{e}}})}_{S,\pm}\boldsymbol{)}
 & = & e^{\mp i \frac{1}{2} \int_{\Gamma}(1-\cos \theta) d\phi}
\nonumber \\
 & & \times U_S (\theta_1,\phi_1)
\ket{\pm} \bra{\pm} U_S^{\dagger} (\theta_0,\phi_0) .
\label{ex.stats.uhl.h}
\end{eqnarray}
Note that the associated holonomies are $g$-independent and satisfy the product relation
\begin{eqnarray}
U_{\textrm{uhl}} \boldsymbol{(} C^{(l,\pm \mu_{\textrm{\scriptsize{e}}})}_{J,\pm}\boldsymbol{)}=
U_{\textrm{uhl}} \boldsymbol{(} C^{(l,\pm \mu_{\textrm{\scriptsize{e}}})}_{L,\pm}\boldsymbol{)}\otimes
U_{\textrm{uhl}} \boldsymbol{(} C^{(l,\pm \mu_{\textrm{\scriptsize{e}}})}_{S,\pm}\boldsymbol{)}.
\label{ex.product relation}
\end{eqnarray}

Next, we compute the Uhlmann holonomy in adiabatic evolution of non-extremal
energy eigenstates characterized by $|\mu| < l+\frac{1}{2}$. The marginal
states
are rank-two density operators obtained by adiabatically evolving the states $\rho_{L,\pm}^{(l,\mu)}$ and $\rho_{S,\pm}^{(l,\mu)}$ under unitaries $U_L$ and $U_S$, respectively. One may solve Eq.~(\ref{eq:ptspherical}) with reference
states $\rho_{L,\pm}^{(l,\mu)}$ for a path $\Gamma$ on the parameter sphere to obtain
the Uhlmann holonomy of the $L$ subsystem,
\begin{eqnarray}
U_{\textrm{uhl}} \boldsymbol{(} C^{(l,\mu)}_{L,\pm}\boldsymbol{)}& = &
U_L (\theta_1,\phi_1) {\bf P} e^{-i \int_{\Gamma} \left( A_{L;\theta}^{(l,\mu)} +
A_{L;\phi}^{(l,\mu)} \right)}
\nonumber \\
 & & \times U_L^{\dagger}(\theta_0,\phi_0).
\label{eq:Uhl.hol.Hy-L}
\end{eqnarray}
Here, we have introduced the vector potential components
\begin{eqnarray}
A_{L;\theta}^{(l,\mu)} & = & \frac{1}{2} w\mathcal{C}^{(l,\mu)} \left( \begin{array}{cc}
0 & -ie^{i\phi} \\
ie^{-i\phi} & 0
\end{array} \right) d\theta ,
\nonumber \\
A_{L; \phi} ^{(l,\mu)} & = & \mu (1-\cos\theta)
\left( \begin{array}{cc}
1 & 0 \\
0 & 1 \end{array} \right) d\phi
\\
 & & + \frac{1}{2}
 \left( \begin{array}{cc}
-1+\cos\theta & w\mathcal{C}^{(l,\mu)} \sin \theta e^{i\phi} \\
w\mathcal{C}^{(l,\mu)}\sin \theta e^{-i\phi} & 1-\cos\theta
\end{array} \right) d\phi .
\nonumber
\end{eqnarray}
The vector potential $A_L^{(l,\mu)}= A_{L;\theta}^{(l,\mu)} +
A_{L;\phi}^{(l,\mu)}$ is expressed in the basis
$\{\ket{\mu-\frac{1}{2}}\ ,\ \ket{\mu+\frac{1}{2}}\}$ and $w=\sqrt{\left( l+\frac{1}{2}
\right)^2 - \mu^2} > 0$.

It is worth noting that the vector potential $A_L^{(l,\mu)}$ exhibits a U(1) part being
proportional to the identity. In the case where $\Gamma$ is a loop, this part gives rise
to the global geometric phase factor $e^{-i\mu \oint_{\Gamma} (1-\cos \theta) d\phi} =
e^{-i\mu \Omega}$, $\Omega$ being the solid angle enclosed by $\Gamma$ on the parameter
sphere $\mathcal{S}^2$.

Similarly, for the $S$ subsystem we have
\begin{eqnarray}
U_{\textrm{uhl}} \boldsymbol{(} C^{(l,\mu)}_{S,\pm}\boldsymbol{)} & = &
U_S (\theta_1,\phi_1) {\bf P} e^{-i \int_{\Gamma} \left( A_{S;\theta}^{(l,\mu)} +
A_{S;\phi}^{(l,\mu)} \right)}
\nonumber \\
 & & \times U_S^{\dagger}(\theta_0,\phi_0)
\label{eq:Uhl.hol.Hy-S}
\end{eqnarray}
with vector potential components
\begin{eqnarray}
A_{S;\theta}^{(l,\mu)} & = &\frac{1}{2} \mathcal{C}^{(l,\mu)}
\left( \begin{array}{cc}
0 & ie^{-i\phi} \\
-ie^{i\phi} & 0
\end{array} \right) d\theta ,
\nonumber \\
A_{S; \phi}^{(l,\mu)}& = &\frac{1}{2}\left( \begin{array}{cc}
1-\cos\theta &
\mathcal{C}^{(l,\mu)} \sin \theta e^{-i\phi} \\
\mathcal{C}^{(l,\mu)} \sin \theta e^{i\phi} &
-1+\cos\theta
\end{array} \right) d\phi
\nonumber \\
\end{eqnarray}
in the $\left\{\ket{+},\ket{-}\right\}$ basis. Note that $A_S^{(l,\mu)}$ does not have
a $U(1)$ part.

Unlike the extremal case, the marginal Uhlmann holonomies are $g$-dependent via the
concurrence $\mathcal{C}^{(l,\mu)}$. Furthermore, there is a dimensional mismatch
between the rank-one holonomy of the $J$ system and the rank-two holonomies of the
$L$ and $S$ subsystems; in general this mismatch implies that there is no path for
which a product rule similar to that in  Eq. (\ref{ex.product relation}) holds.

It is noticed  that when $\alpha^{(l,\mu)} \neq 0,\pi$, i.e., when the
concurrence $\mathcal{C}^{(l,\mu)}$ is non-zero, the vector potentials contain non-Abelian
components. In other words, the non-Abelian nature of the subsystem holonomies is due to
entanglement. This is analogous to the L\'{e}vay geometric phase defined for two-qubit systems
\cite{levay04a}, which is a path-dependent unit quaternion that may find realization in
two-particle interferometry \cite{johansson11}. The holonomy group associated with this
geometric phase becomes, just as the Uhlmann holonomy of the subsystems, Abelian in the
product state case.

In the ``classical'' limit characterized by $l \rightarrow \infty$ and $|\mu| =
{\textrm{O}} (l)$, $\mathcal{C}^{(l,\mu)}$ vanishes as $\alpha^{(l,\mu)}$ tends to
zero. Thus, the $L$ and $S$ holonomies turn Abelian in this limit. We further see
that for cyclic evolution in the $l/|\mu|,l/|g| \rightarrow \infty$ limit,
$\mathcal{C}^{(l,\mu)}$ tends to its maximum value $1$, i.e., the energy eigenvector
becomes maximally entangled, and the holonomies turn into the transpose of the
Wilczek-Zee holonomy for nuclear quadrupole resonance setup discussed in
Ref. \cite{zee88} and experimentally studied in Refs. \cite{tycko87,zwanziger90}.

\subsection{Figure-8 Curve}
\begin{figure}[h]
\centering
\includegraphics[scale=0.50]{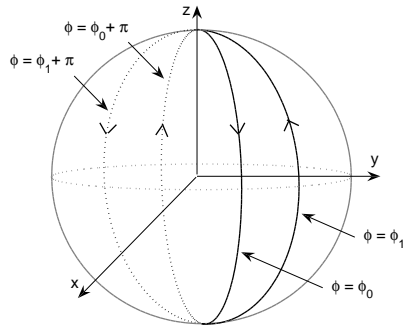}
\caption{Figure-8 loop on the parameter sphere $\mathcal{S}^2$ of magnetic field
directions. The Abelian geometric phase of the whole system along this
path is zero because the net enclosed area is zero. In contrast, the phases of the
Wilson loop variable associated with the Uhlmann holonomy for the $L$ and $S$ are in
general nonzero.}
\label{fig:Figure8}
\end{figure}

In order to investigate the consequences of the non-Abelian structure of the
Uhlmann holonomies of the $L$ and $S$ subsystems, we consider here the class of
``figure-8'' loops on the parameter sphere of magnetic field directions shown in
Fig. \ref{fig:Figure8}, for which the holonomies can be calculated explicitly.
These loops are chosen to enclose no net area. Since any Abelian geometric phase
of a angular momentum system is proportional to the area enclosed on the parameter
sphere \cite{berry84}, any such phase must vanish for the figure-8 loops; a fact
that has been used to demonstrate the Abelian nature of the Berry phase experimentally
by using nuclear magnetic resonance techniques \cite{appelt95}. Similarly, the
mixed-state geometric phases \cite{sjoqvist00a} and the Berry phases in the $LS$
system are all zero due to their Abelian nature. In contrast, we show that the
Uhlmann phases of the subsystems along this class of loops are in general non-zero,
which is a clear signature of the non-Abelian structure of the underlying Uhlmann
holonomy.

One can parametrize the loop $\Gamma = \Gamma'' \circ \ \Gamma'$ in Fig.
\ref{fig:Figure8} as
\begin{eqnarray}
\Gamma'(\theta_t,\phi_t) = \left\{ \begin{array}{l}
\Gamma'_0:(\theta'_0(t)  \in [0 ,\pi), \phi_0) \\
\Gamma'_1:(\pi,\phi'_0(t) \in [\phi_0 ,\phi_1]) \\
\Gamma'_2:(\theta'_1(t) \in (\pi , 0), \phi_1 )\\
\Gamma'_3: (0, \phi'_1(t) \in [\phi_1 ,\phi_1+\pi])
\end{array} \right.
\label{orangeslice}
\end{eqnarray}
and
\begin{eqnarray}
\Gamma''(\theta_t,\phi_t) = \left\{ \begin{array}{l}
\Gamma''_0: (\theta''_0(t)  \in [0 ,\pi), \phi_1+\pi) \\
\Gamma''_1: (\pi,\phi''_0(t) \in [\phi_1+\pi ,\phi_0+\pi] \\
\Gamma''_2: (\theta''_1(t) \in (\pi , 0), \phi_0+\pi)\\
\Gamma''_3: (0, \phi''_1(t) \in [\phi_0+\pi ,\phi_0])
\end{array} \right.
\end{eqnarray}
with $\{ \theta'_0 (t) , \ldots , \phi''_1 (t) \}$ being a time ordered set
of smooth functions.

For the extremal subspaces $\mu = \pm \mu_{\textrm{\scriptsize{e}}}$, the holonomies
of the $L,S,$ and $J$ systems are trivial in the sense that
$U_{\textrm{uhl}} \boldsymbol{(} C^{(l,\mu)}_{X,\pm}\boldsymbol{)}$, $X=L,S,J$,
become projection operators. This follows from the Abelian nature of the extremal
states and from the fact that the net area vanishes for $\Gamma$.

On the other hand, in the case where $|\mu| <l+\frac{1}{2}$, the holonomies
turn non-Abelian and the corresponding Uhlmann phases might be non-zero.
We demonstrate this in detail for the $L$ subsystem.

By using that $U_L(0,\phi) = \hat{1}$, integration along $\Gamma'$ and $\Gamma''$
yields
\begin{eqnarray}
U_{\textrm{uhl}} \boldsymbol{(} C'^{(l,\mu)}_{L,\pm}\boldsymbol{)}& = &
e^{-i2\mu(\phi_1-\phi_0)}\left( \begin{array}{cc}
a_L& b_L\\
-b_L^*& a_L^*
\end{array} \right) ,
\nonumber \\
U_{\textrm{uhl}} \boldsymbol{(} C''^{(l,\mu)}_{L,\pm}\boldsymbol{)} & = &
e^{i2\mu(\phi_1-\phi_0)} \left( \begin{array}{cc}
a_L^*& b_L \\
-b_L^*& a_L
\end{array} \right) .
\label{orange.s.Uhl}
\end{eqnarray}
Here,
\begin{eqnarray}
a_L & = & e^{i(\phi_1-\phi_0)} \cos^2 \left( \frac{\chi_L}{2} \right) +
\sin^2 \left( \frac{\chi_L}{2} \right) ,
\nonumber \\
b_L & = & \cos \left( \frac{\chi_L}{2} \right) \sin \left( \frac{\chi_L}{2} \right)
\left( - e^{i\phi_1} + e^{i\phi_0} \right) ,
\label{ab-l system}
\end{eqnarray}
where $\chi_L=w\mathcal{C}^{(l,\mu)}\pi$.
Consequently
\begin{eqnarray}
U_{\textrm{uhl}} \boldsymbol{(} C^{(l,\mu)}_{L,\pm}\boldsymbol{)} = \left( \begin{array}{cc}
\left|a_L\right|^2-\left|b_L\right|^2& 2a_L^*b_L \\
-2a_Lb_L^*& \left|a_L\right|^2-\left|b_L\right|^2
\end{array} \right) ,
\label{f8Uhl}
\end{eqnarray}
where $C^{(l,\mu)}_{L,\pm} = C''^{(l,\mu)}_{L,\pm} \circ C'^{(l,\mu)}_{L,\pm}$.

We may interpret the Uhlmann holonomy in Eq. (\ref{f8Uhl}) as a rotation in an abstract
space defined by the two states $\ket{l,\mu \pm \frac{1}{2}}$. Explicitly, if we let
$\frac{a_L}{b_L} = \tan \frac{\eta}{2} e^{-i\kappa}$, and introduce the effective
Pauli operators $\sigma_L^x = \ket{l,\mu + \frac{1}{2}} \bra{l,\mu - \frac{1}{2}} +
{\textrm{h.c}}$, $\sigma_L^y = -i \ket{l,\mu + \frac{1}{2}} \bra{l,\mu - \frac{1}{2}} +
{\textrm{h.c}}$, and $\sigma_L^z = \ket{l,\mu + \frac{1}{2}} \bra{l,\mu + \frac{1}{2}} -
\ket{l,\mu - \frac{1}{2}} \bra{l,\mu - \frac{1}{2}}$, defining an internal $xyz$ coordinate
system, the Uhlmann holonomy in  Eq. (\ref{f8Uhl}) can be viewed as a rotation by an angle
$\kappa$ around an axis in the $xy$ plane making an angle $\eta$ with the $x$ axis in the
internal space.

By applying the definition in Eq. (\ref{Uhlmannphase}) to the Uhlmann holonomy given by
Eq. (\ref{f8Uhl}), we obtain the Uhlmann phase as
\begin{eqnarray}
\gamma_{L;\pm}^{(l,\mu)} = \arg \left[ \Tr \left(U_{\textrm{uhl}}
\boldsymbol{(} C^{(l,\mu)}_{L,\pm}\boldsymbol{)} \right) \right] =
\arg \xi ,
\end{eqnarray}
where
\begin{eqnarray}
\xi = 2[\cos \left( \phi_1 - \phi_0 \right)\sin^2 \left(\chi_L \right)
+ \cos^2 \left( \chi_L\right)] .
\label{up}
\end{eqnarray}
Hence,
\begin{eqnarray}
\gamma_{L;\pm}^{(l,\mu)} =
\left\{ \begin{array}{l}
0, \ \ \xi > 0 \\
\pi , \ \ \xi < 0 \\
{\textrm{undefined}}, \ \ \xi = 0 .
\end{array} \right.
\label{f8uhlphase}
\end{eqnarray}
The points in the space $(\phi_1-\phi_0,\chi_L \propto \mathcal{C}^{(l,\mu)})$ where $\xi$
vanishes form a nodal line along which the Uhlmann phase $\gamma_{L;\pm}^{(l,\mu)}$ is
undefined. The points  along this line are analogous to the nodal points found in Ref.
\cite{sjoqvist05} in the case of the mixed-state geometric phase for this system.

\begin{figure}[h]
\centering
\includegraphics[scale=0.50]{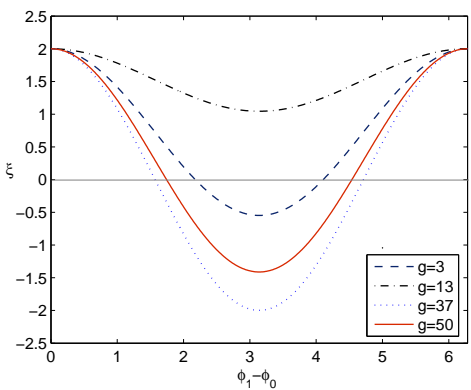}
\caption{(Color online) $\xi$ defined in Eq. (\ref{up}) as a function of  $\phi_1 - \phi_0$,
for $l=3$, $\mu=\frac{3}{2}$, and the coupling strengths $g=3,13,37,50$. The corresponding
phase $\gamma_{L;\pm}^{(l,\mu)}$ of the Wilson loop variable associated with the Uhlmann
holonomy of the $L$ subsystem is given $\arg \xi$.}
\label{fig:Hydrogen-like1}
\end{figure}

In Fig.~\ref{fig:Hydrogen-like1} we have plotted $\xi$, as a function of $\phi_1-\phi_0$
for $l=3$, $\mu = \frac{3}{2}$, and $g=3,13,37,50$. The figure shows that $\xi$ can in
fact satisfy each of the three possible conditions displayed in Eq.~(\ref{f8uhlphase}).
In other words, there exist figure-8 loops for which the Uhlmann phase
$\gamma_{L;\pm}^{(l,\mu)}$ is $\pi$, in contrast to the corresponding mixed-state
geometric phases that always vanish for such loops. This result is due to the
non-Abelian nature of the underlying Uhlmann holonomy for non-extremal states.
The existence of non-zero $\gamma_{L;\pm}^{(l,\mu)}$ is furthermore related to
entanglement in a non-trivial way: one can show that a non-zero $\gamma_{L;\pm}^{(l,\mu)}$
may occur only if $\frac{1}{4\sqrt{(l+\frac{1}{2})^2-\mu^2}} < \mathcal{C}^{(l,\mu)} <
\frac{3}{4\sqrt{(l+\frac{1}{2})^2-\mu^2}}$.

One may verify that the phase $\gamma_{S;\pm}^{(l,\mu)}$ of the Wilson loop variable
associated with the Uhlmann holonomy of the $S$ subsystem may similarly be $\pi$ for
certain figure-8 loops, while $0$ or undefined for other loops. Note in particular that
the necessary condition on concurrence for a non-zero $\gamma_{S;\pm}^{(l,\mu)}$ now
reads $\frac{1}{4} < \mathcal{C}^{(l,\mu)} < \frac{3}{4}$, which is different from the
above condition for the $L$ subsystem. Thus, there may be energy eigenstates that allows
for a non-zero $\gamma_{S;\pm}^{(l,\mu)}$ while $\gamma_{L;\pm}^{(l,\mu)}$ must be zero
and viceversa. In fact, if $\left( l+\frac{1}{2} \right)^2 -\mu^2 > 9$, then only one
of the two Uhlmann phases can be $\pi$ for any given loop on the parameter sphere.

\subsection{Additivity}
In this section, we explore the relation between the Uhlmann phases of the whole system and
the subsystems. We restrict to cyclic evolutions, for which the Uhlmann holonomy takes the
Wilson loop form.

For $\mu = \pm \mu_{\textrm{\scriptsize{e}}}$, Eq. (\ref{ex.product relation})
yields the sum rule  $\gamma_{J;\pm}^{\pm \mu_{\textrm{\scriptsize{e}}}}=
\gamma_{L;\pm}^{\pm \mu_{\textrm{\scriptsize{e}}}}+\gamma_{S;\pm}^{\pm \mu_{\textrm{\scriptsize{e}}}}$
for the corresponding Uhlmann phases. This confirms the expected sum rule for the geometric phases of
product states.

For $|\mu| <l+\frac{1}{2}$, we put $A_L^{(l,\mu)} = A_{L;{\textrm{U(1)}}}^{(l,\mu)} +
A_{L;{\textrm{SU(2)}}}^{(l,\mu)}$, where $A_{L;{\textrm{U(1)}}}^{(l,\mu)} = \mu (1-\cos \theta)d\phi$,
and obtain
\begin{eqnarray}
\gamma_{J;\pm}^{(l,\mu)}& = & \gamma_{L;\pm}^{(l,\mu)} + \gamma_{S;\pm}^{(l,\mu)}
\nonumber\\
 & & -\arg\left[ \Tr \left({\bf P} e^{-i \oint_{\Gamma} A_{L;{\textrm{SU(2)}}}^{(l,\mu)}}
\right)\Tr \left({\bf P} e^{-i \oint_{\Gamma} A_{S;{\textrm{SU(2)}}}^{(l,\mu)}}\right)\right]
\nonumber\\
 & \equiv & \gamma_{L;\pm}^{(l,\mu)} +\gamma_{S;\pm}^{(l,\mu)}+ \Delta \gamma^{(l,\mu)} .
\end{eqnarray}
Since the trace of an SU(2) matrix is real, the deviation $\Delta \gamma^{(l,\mu)}$ from
the sum rule $\gamma_{J;\pm}^{(l,\mu)} =  \gamma_{L;\pm}^{(l,\mu)} + \gamma_{S;\pm}^{(l,\mu)}$
can only be $\pi$ for a cyclic evolution in this model system. Similar to the
nodal points found in Ref.~ \cite{sjoqvist05} for the mixed-state geometric
phases \cite{sjoqvist00a} of the $L$ and $S$ subsystems, there
exist loops for which either $\Tr \left({\bf P} e^{-i \oint_{\Gamma}
A_{L;{\textrm{SU(2)}}}^{(l,\mu)}} \right)$ or $\Tr \left({\bf P}
e^{-i \oint_{\Gamma} A_{S;{\textrm{SU(2)}}}^{(l,\mu)}}\right)$ vanish so
that $\Delta \gamma^{(l,\mu)}$ becomes undefined. These loops are nodal points of
$\gamma_{L;\pm}^{(l,\mu)}$ or $\gamma_{S;\pm}^{(l,\mu)}$, respectively.

We demonstrate that $\Delta \gamma^{(l,\mu)}$ can be zero, non-zero, or undefined for the
``orange slice'' loop defined in Eq. (\ref{orangeslice}), which connects the two
poles on the parameter sphere twice along geodesics at $\phi_0$ and $\phi_1$.

\begin{figure}[h]
\centering
\includegraphics[scale=0.50]{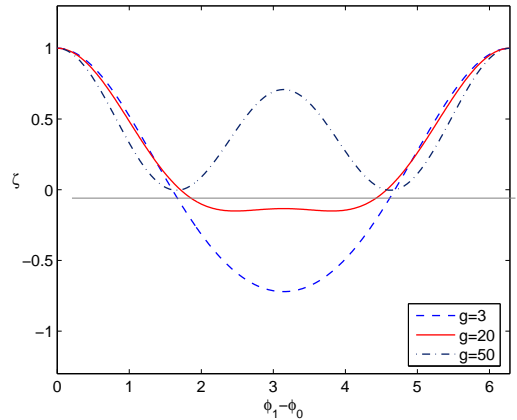}
\caption{(Color online) $\zeta$ defined in Eq. (\ref{eq:zeta}) as a function of $\phi_1 - \phi_0$
for $l=2$, $\mu=\frac{1}{2}$, and $g = 3, 20, 50$. The difference $\Delta \gamma^{(l,\mu)} =
\gamma_{J;\pm}^{(l,\mu)} - \gamma_{L;\pm}^{(l,\mu)} - \gamma_{S;\pm}^{(l,\mu)}$ of the phases
of the corresponding Wilson loop variables is given by $-\arg \zeta$.}
\label{fig:Hydrogen-like}
\end{figure}

The holonomy of the $L$ subsystem for the orange slice loop $\Gamma'$ can be found in
Eq. (\ref{orange.s.Uhl}). For the $S$ subsystem, we obtain
\begin{eqnarray}
U_{\textrm{uhl}} \boldsymbol{(} C'^{(l,\mu)}_{S,\pm}\boldsymbol{)}& = &
\left( \begin{array}{cc}
a_S^*& -b_S^*\\
b_S& a_S
\end{array} \right) ,
\end{eqnarray}
where $a_S$ and $b_S$ are obtained from $a_L$ and $b_L$ in Eq.~(\ref{ab-l system})
by replacing $\chi_L$ with $\chi_S=\mathcal{C}^{(l,\mu)}\pi$. The corresponding Uhlmann
phases read
\begin{eqnarray}
\gamma_{J;\pm}^{(l,\mu)} & = & -\mu 2(\phi_1-\phi_0) ,
\nonumber \\
\gamma_{L;\pm}^{(l,\mu)} & = & -\mu 2(\phi_1-\phi_0)\nonumber \\
&&+ \arg \left[ \cos^2 \left( \frac{\chi_L}{2} \right)
\cos \left( \phi_1 - \phi_0 \right) +
\sin^2 \left( \frac{\chi_L}{2} \right) \right] ,
\nonumber \\
\gamma_{S;\pm}^{(l,\mu)} & = & \arg \left[ \cos^2 \left( \frac{\chi_S}{2} \right)
\cos \left( \phi_1 - \phi_0 \right) +
\sin^2 \left( \frac{\chi_S}{2} \right) \right].\nonumber \\
\end{eqnarray}
Thus, $\Delta \gamma^{(l,\mu)} \equiv -\arg \zeta$, where
\begin{eqnarray}
\zeta & = & \left[ \cos^2 \left( \frac{\chi_L}{2} \right)
\cos \left( \phi_1 - \phi_0 \right) +
\sin^2 \left( \frac{\chi_L}{2} \right) \right]
\nonumber \\
 & & \times \left[ \cos^2 \left( \frac{\chi_S}{2} \right)
\cos \left( \phi_1 - \phi_0 \right) +
\sin^2 \left( \frac{\chi_S}{2} \right) \right] .
\label{eq:zeta}
\end{eqnarray}
$\Delta \gamma^{(l,\mu)}$ is $0$ if $\zeta > 0$, $\pi$ if $\zeta < 0$, and undefined if $\zeta = 0$.
All these three cases are visible in Fig.~(\ref{fig:Hydrogen-like}), in which we have plotted
$\zeta$ as a function of $\phi_1 - \phi_0$ for fixed $l$, $\mu$, and $g = 3, 20, 50$.
Figure \ref{fig:Hydrogen-like} confirms that there are some loops on the parameter
sphere for which $\Delta \gamma^{(l,\mu)} \neq 0$, and therefore $\gamma_{J;\pm}^{(l,\mu)}
\neq \gamma_{L;\pm}^{(l,\mu)} + \gamma_{S;\pm}^{(l,\mu)}$.

To compare with the mixed state geometric phase proposed in Ref. \cite{sjoqvist00a}, we
first consider the spectral decomposition $\{ p_k,\ket{\psi_k} \}$ of an arbitrary non-degenerate
density operator $\rho$ that undergoes cyclic unitary evolution. The resulting mixed
state geometric phase $\beta$ reads $\beta = \arg \left( \sum_k p_k e^{i\beta_k} \right)$,
where $\beta_k$ is the cyclic pure state geometric phase of $\ket{\psi_k}$. Using this expression,
the mixed-state geometric phases $\beta(C_{L,\pm}^{(l;\mu)})$ and $\beta(C_{S,\pm}^{(l;\mu)})$
of the $L$ and $S$ subsystems read \cite{sjoqvist05}
\begin{eqnarray}
\beta(C_{L,\pm}^{(l;\mu)}) & = &
-\mu \Omega \pm \arctan \left( \cos \alpha^{(l;\mu)}
\tan \frac{\Omega}{2} \right) ,
\nonumber \\
\beta(C_{S,\pm}^{(l;\mu)}) & = &
\mp \arctan \left( \cos \alpha^{(l;\mu)}
\tan \frac{\Omega}{2} \right) ,
\label{eq:marginalgp}
\end{eqnarray}
which implies the sum rule $\beta(C_{L,\pm}^{(l;\mu)}) + \beta(C_{S,\pm}^{(l;\mu)}) =
-\mu \Omega = \beta(C_{J,\pm}^{(l;\mu)})=\gamma_{J;\pm}^{(l,\mu)}$. Thus, the sum rule
for $\beta$ is satisfied for any loop $\Gamma$ on the parameter sphere.
This again demonstrates the difference between the phase of the Wilson loop variable
associated with the Uhlmann holonomy and the mixed-state geometric phase defined in
Ref.~\cite{sjoqvist00a}.

We note that, the alternative Uhlmann phase $\varphi_{\textrm{uhl}}$ defined in Eq.
\ref{Uhlmannphase0} is in general non-zero for the figure-8 path in Fig. \ref{fig:Figure8},
as a consequence of the non-Abelian nature of the underlying Uhlmann holonomy. On the other
hand, the corresponding deviation $\Delta \varphi_{\textrm{uhl}}^{(l,\mu)}$ from the sum rule is
no longer restricted to $0$ or $\pi$, but can take any real value for cyclic evolutions. Thus,
in general there is a difference between  $\varphi_{\textrm{uhl}}$ and $\gamma$, and they both
differ, in the sense of the sum rule, from the mixed-state geometric phase in Ref.~\cite{sjoqvist00a}.

\section{Conclusions}
Uhlmann's quantum holonomy along density operators is a concept that allows
for studies of geometric phases of general quantum states undergoing arbitrary
quantum evolutions. Its relevance to various aspects of physics have been
demonstrated in the past, such as for open system evolution \cite{tidstrom03},
interferometry \cite{aberg07}, many-body quantum systems \cite{paunkovic08},
as well as for Yang-Mills theory \cite{dittmann98} and Thomas precession in special
relativity \cite{levay04b}. These works have been triggered in part by the need
to test the conjectured resilience of holonomic quantum gates for quantum computation
to various kinds of errors, such as noise and decoherence. Recently, a first experimental
test of the related Uhlmann geometric phase, utilizing nuclear magnetic resonance
techniques, has been carried out \cite{du07}.

Here, we have analyzed Uhlmann's quantum holonomy by considering a physical model
system in which the Uhlmann holonomies for $LS$-coupled hydrogen-like atoms in a slowly
rotating magnetic field have been computed. We have shown that the holonomy of the
total angular momentum has Abelian structure. Furthermore, its corresponding phase
is exactly the associated standard geometric phase\cite{mukunda93} for open or closed
paths on the parameter sphere of magnetic field directions. For the holonomies of the
$L$ and $S$ subsystems, we have shown that, in analogy with the L\'{e}vay geometric phase
defined for two-qubit systems \cite{levay04a}, depending on whether the energy eigenstate
of the whole system is a product state or an entangled state, the  corresponding holonomies
are Abelian or non-Abelian, respectively. In the case of entangled states, there is an
explicit dependence of the gauge field vector potential on the concurrence \cite{wootters98},
which interpolates the standard Abelian (Berry) and non-Abelian (Wilczek-Zee) cases.
In other words, our analysis demonstrates that the rich geometrical nature of the Uhlmann
holonomy incorporates as a limiting case the  Wilczek-Zee holonomy, which is characterized
by maximum quantum entanglement between the $L$ and $S$ subsystems.

In the analysis of the phase of the Wilson loop variable associated with the Uhlmann holonomy,
we have pointed out that this phase, unlike the mixed-state geometric phase \cite{sjoqvist00a},
possesses a non-Abelian structure and may therefore be non-zero even for loops on the parameter
sphere that enclose no net area. We have also elucidated that the phases of the Wilson loop
variables of the corresponding Uhlmann holonomies for the $L$ and $S$ subsystems add up to that
of the whole system for specific paths; for other paths the sum may differ by $\pi$ from the
Berry phase of the whole system.

Furthermore, we would like to point out that previous theoretical
\cite{slater02,ericsson03,rezakhani06} and experimental \cite{du07} work analyzing the
relation between the Uhlmann holonomy and the mixed-state geometric phase all employ a
notion of Uhlmann phase [see Eq.~(\ref{Uhlmannphase0})] which differs both conceptually
and numerically from the phase of the Wilson loop variable associated with the Uhlmann
holonomy, as defined in  Eq.~(\ref{Uhlmannphase}). This alternative concept is the phase
of the Hilbert-Schmidt overlap between the initial and the parallel transported final
Uhlmann amplitude. In Ref.~\cite{du07} it was also pointed out that other definitions of
mixed-state geometric phases, which differ both from the geometric phase considered in
Ref.~\cite{sjoqvist00a} and the (Hilbert-Schmidt) Uhlmann phase are in principle possible
and experimentally relevant. It is important that these different definitions of mixed-state
geometric phases be thoroughly investigated and compared with the phases associated to the
Uhlmann holonomy in model systems, where exact or computationally feasible solutions exist.
Our paper is a contribution in this direction.

The results for the phase of the Wilson loop variable suggest it would be of interest to
test the relation between this phase and various mixed-state geometric phases experimentally.
This would further improve our understanding of the relation between the Uhlmann holonomy
\cite{uhlmann86} and the mixed-state geometric phase \cite{sjoqvist00a}. It would also help
shed light on which of these phases is the most robust and, at the same time, the most accessible
experimentally and the most amenable to external manipulation.

\section*{ACKNOWLEDGMENTS}
V.A.M. and C.M.C. acknowledge support of the Faculty of Natural Sciences at Linnaeus University,
and the Swedish Research Council under Grant Numbers: 621-2007-5019 and 621-2010-3761.
E.S. acknowledges support from the National Research Foundation and the Ministry of Education
(Singapore).


\begin{thebibliography}{99}
\bibitem{berry84} M. V. Berry,
Proc. R. Soc. London Ser. A {\bf 392}, 45 (1984).
\bibitem{wilczek84} F. Wilczek and A. Zee,
Phys. Rev. Lett. {\bf 52}, 2111 (1984).
\bibitem{moody86} J. Moody, A. Shapere, and F. Wilczek,
Phys. Rev. Lett. {\bf 56}, 893 (1986).
\bibitem{tycko87} R. Tycko,
Phys. Rev. Lett. {\bf 58}, 2281 (1987).
\bibitem{arovas98} D. P. Arovas and Y. Lyanda-Geller,
Phys. Rev. B {\bf 57}, 12302 (1998).
\bibitem{unanyan99} R. G. Unanyan, B. W. Shore, and K. Bergmann,
Phys. Rev. A {\bf 59}, 2910 (1999).
\bibitem{pachos00} J. Pachos and S. Chountasis,
Phys. Rev. A {\bf 62}, 052318 (2000).
\bibitem{faoro03} L. Faoro, J. Siewert, and R. Fazio,
Phys. Rev. Lett. {\bf 90}, 028301 (2003).
\bibitem{zanardi99} P. Zanardi and M. Rasetti,
Phys. Lett. A {\bf 264}, 94 (1999).
\bibitem{uhlmann86} A. Uhlmann,
Rep. Math. Phys. {\bf 24}, 229 (1986).
\bibitem{sjoqvist05} E. Sj\"oqvist, X. X. Yi, and J. {\AA}berg,
Phys. Rev. A {\bf 72}, 054101 (2005).
\bibitem{oh08} S. Oh, Z. Huang, U. Peshkin, and S. Kais,
Phys. Rev. A {\bf 78}, 062106 (2008).
\bibitem{sjoqvist00a}  E. Sj\"{o}qvist, A. K. Pati, A. Ekert,
J. S. Anandan, M. Ericsson, D. K. L. Oi, and V. Vedral,
Phys. Rev. Lett. {\bf 85}, 2845 (2000).
\bibitem{remark1} An operator $O$ is a partial isometry if $OO^{\dagger}$ and
$O^{\dagger}O$ are projection operators. In particular, the phase factor $V_t$
should be a partial isometry satisfying $\sqrt{\rho_t} V_t V_t^{\dagger} \sqrt{\rho_t} =
\rho_t$.
\bibitem{hubner93} M. H\"ubner,
Phys. Lett. A {\bf 179}, 226 (1993).
\bibitem{mukunda93} N. Mukunda and R. Simon,
Ann. Phys. (N.Y.) {\bf 228}, 205 (1993).
\bibitem{slater02} P. B. Slater,
Lett. Math. Phys. {\bf 60}, 123 (2002).
\bibitem{ericsson03} M. Ericsson, A. K. Pati, E. Sj\"{o}qvist,
J. Br\"{a}nnlund and D. K. L. Oi,
Phys. Rev. Lett. {\bf 91}, 090405 (2003).
\bibitem{rezakhani06} A. T. Rezakhani and P. Zanardi,
Phys. Rev. A {\bf 73}, 012107 (2006).
\bibitem{du07} J. Zhu, M. Shi, V. Vedral, X. Peng, D. Suter, and J. Du,
Europhys. Lett. {\bf 94}, 20007 (2011).
\bibitem{wu75} T. T. Wu and C. N. Yang,
Phys. Rev. D {\bf 12}, 3845 (1975).
\bibitem{wootters98} W. K. Wootters,
Phys. Rev. Lett. {\bf 80}, 2245 (1998).
\bibitem{levay04a} P. L\'{e}vay,
J. Phys. A: Math. Gen. {\bf 37}, 1821 (2004).
\bibitem{johansson11} M. Johansson, M. Ericsson, K. Singh, E. Sj\"oqvist, and M. S. Williamson,
J. Phys. A: Math. Theor. {\bf 44}, 145301 (2011).
\bibitem{zee88} A. Zee,
Phys. Rev. A {\bf 38}, 1 (1988).
\bibitem{zwanziger90} J. W. Zwanziger, M. Koenig, and A. Pines,
Phys. Rev. A {\bf 42}, 3107 (1990)
\bibitem{appelt95} S. Appelt, G. W\"ackerle, and M. Mehring,
Phys. Lett. A {\bf 204}, 210 (1995).
\bibitem{tidstrom03} J. Tidstr\"om and E. Sj\"oqvist,
Phys. Rev. A {\bf 67}, 032110 (2003).
\bibitem{aberg07} J. {\AA}berg, D. Kult, E. Sj\"oqvist, and D. K. L. Oi,
Phys. Rev. A {\bf 75}, 032106 (2007).
\bibitem{paunkovic08} N. Paunkovic and V. R. Vieira,
Phys. Rev. E {\bf 77}, 011129 (2008).
\bibitem{dittmann98} J. Dittmann,
Lett. Math. Phys. {\bf 46}, 281 (1998).
\bibitem{levay04b} P. L\'{e}vay,
J. Phys. A: Math. Gen. {\bf 37}, 4593 (2004).
\end{thebibliography}
\end{document}